# New models for the rapid evolution of the central star of the Stingray Nebula

T.M. Lawlor

Pennsylvania State University, Brandywine, Department of Physics, Media, PA 19063 USA

**ABSTRACT**

We present stellar evolution calculations from the Asymptotic Giant Branch (AGB) to the Planetary Nebula (PN) phase for models of initial mass 1.2 $M_\odot$ and 2.0 $M_\odot$ that experience a Late Thermal Pulse (LTP), a helium shell flash that occurs following the AGB and causes a rapid looping evolution between the AGB and PN phase. We use these models to make comparisons to the central star of the Stingray Nebula, V839 Ara (SAO 244567). The central star has been observed to be rapidly evolving (heating) over the last 50 to 60 years and rapidly dimming over the past 20 - 30 years. It has been reported to belong to the youngest known planetary nebula, now rapidly fading in brightness. In this paper we show that the observed timescales, sudden dimming, and increasing Log(g), can all be explained by LTP models of a specific variety. We provide a possible explanation for the nebular ionization, the 1980's sudden mass loss episode, the sudden decline in mass loss, and the nebular recombination and fading.

## INTRODUCTION

The central star of the Stingray Nebula, V839 Ara (SAO244567) has been observed to be evolving on a very rapid timescale. It has increased in temperature by some 30,000K over a mere 30 years while also decreasing in brightness dramatically since the late 1980s (Parthasarathy, 2000, Schaefer et al. 2021 – SBS21 hereafter). The central star, also known as Hen 1357, was described as being a type B or A giant in 1950 and a type B1 supergiant in 1971 (Henize, 1976). However, in 1990 the spectrum was more like that of a CSPN, and was declared at that time to be the central star of the youngest known planetary nebula (Parthasarathy, 1993; Bobrowsky, 1994).

Bobrowsky (1994) estimated the central star mass to be 0.59 $M_\odot$, and Parthasarathy (2000) estimated 0.6 $M_\odot$. Its distance was initially estimated to be 5.6 kpc (Kozok, 1985) and it was described as being at high galactic latitude by Parthasarathy and Pottasch, (1989). They also estimated a nebula expansion age of 2650 years. Later, Reindl et al. (2014, RNDL14 hereafter) estimated the distance to be 1.6 kpc, concluding only 1000 years for a nebula expansion age. In the same work, RNDL14 estimate the central stellar mass to be as low as 0.35 $M_\odot$ and as high as 0.5 $M_\odot$, however with additional observations Reindl et al. (2017, RNDL17 hereafter) adjusted their estimate for mass to 0.53 $M_\odot$ - 0.56 $M_\odot$. RNDL17 announced that the central star has begun to cool since 2002 and brighten beginning after 2006. To the contrary, SBS21 claim no such cooling, but rather a heating with no brightening - something they admit they "do not understand." They also report that after being ionized by an unknown mechanism and experiencing a mass loss episode in the 1980's, the nebula has dimmed drastically between 2006 and 2016. This was also investigated by Balick et al. (2021) who indicate that "the nebula probably responded very quickly to sudden changes in the ionizing photon flux" during the 1980's.

The cause of V839 Ara's rapid evolution has been proposed to be either due to binary interaction or a helium pulse of some variety, along with some less likely scenarios (Schaefer & Edwards, 2015, SE15 hereafter, RNDL14). RNDL14 suggest that a late thermal pulse is a good candidate to explain the star's rapid evolution, but note that concrete evolutionary models that match this star are missing. RNDL17 make a more firm affirmation that the most likely scenario is an LTP, though still found some short comings due to a dearth of published LTP models. SBS21 expressed some doubts on this determination but did not rule it out. Thus, in this paper we present a new set of late thermal pulse models tailored for this object, and use them for comparisons to observations.

Post-AGB helium flashes of all types have been modeled for some time (Paczynski 1971; Renzini 1979; Schönberner 1979; Iben et al. 1984; Wood & Faulkner 1986; Vassiliadis & Wood 1994; Iben, Tutukov & Yungelson 1996; Herwig et al. 1999; Blocker 2001; Lawlor & MacDonald 2003 (LM03 hereafter); Lawlor and MacDonald 2006). LTP objects have been modelled but discussed less so than very late thermal pulse (VLTP) objects, called 'Born-again' objects early on because the helium flash occurs on the white dwarf cooling track (Iben 1985; Iben & MacDonald 1995). An early first example of LTP 'loops' (so far as the author can tell) was modelled in a short paper by Schwartzschild and Härm (1970) and their consequences for nebulae were discussed in Tylenda, 1979. Blocker (2001) contains an excellent review of such flashes in the context of attempting to explain the origin of Wolf-Rayett central stars and PG 1159 stars by invoking convective overshoot. Lawlor & MacDonald (2006) also make comparisons to PG 1159 stars, and include LTP models in their paper focused on VLTP objects and helium rich white dwarfs. First to coin the designations, Blöcker and Schönberner (1997) and Blöcker (2001) divide the types of helium flashes into three types: A final AGB thermal pulse (AFTP) that occurs while still on the TPAGB with a hydrogen envelope mass around 0.01 $M_\odot$; an LTP, which occurs while evolving at constant luminosity from the AGB to the PNe with an envelope hydrogen mass $\approx 10^{-4}$ $M_\odot$; and a VLTP which occurs after the model star has already entered the white dwarf cooling track with only a thin hydrogen envelope. Lawlor & MacDonald (2006) adopt these designations but divide them further into ten types depending on the metallicity of the star and whether the model departs the AGB burning hydrogen or helium. Pertinent to this paper, they designate LTP models into type III and IV to distinguish at what temperature the late thermal pulse ignites. Type IV occurs at $\text{Log}(T_{\text{eff}}) \lesssim 4.47$ while type III occurs above this temperature. This temperature is approximately the temperature required to ionize a surrounding planetary nebula. They also provide a unitless parameter related to the helium layer mass as a criterion for post-AGB flash behaviour.

In section 2 of this paper we describe our code and how we construct our models, and provide a representative description of LTP model behaviour for a range of helium layer masses. In section 3 we describe observations for V839 Ara taken from a number of sources that we use for model-observation comparisons and we explain our choice of models for the comparisons. We make a variety of model-observation comparisons including HR diagram evolution tracks, the Log(g)-Log($T_{eff}$) plane, and to evolutionary time scales. We discuss what we can conclude from our comparisons and summarize and provide discussion in section 4.

## 2 LATE THERMAL PULSE MODELS

In this section we describe our stellar evolution code and how we construct our pertinent grid of late thermal pulse models and we describe our choice of models for comparison. We show representative LTP evolution tracks for a general overview of LTP behaviour over a range of helium masses at the end of the AGB. We also highlight the variation in behaviour which depends on at what temperature the helium flash ignites. This is a critical attribute for explaining the behaviour of V839 Ara.

### 2.1 Stellar Evolution Code

Our late thermal pulse models are calculated using the evolution code BRAHMA, described in numerous contexts including post-AGB helium pulses (LM03, Lawlor, 2005; Lawlor & MacDonald, 2006), Brown Dwarfs (Mullan & MacDonald, 2010), and very low metal and Population III stars (MacDonald et al., 2013; Lawlor et al.; 2015). Thus we offer a brief summary of it here and review details pertinent to this work. The code is a distant decedent of the widely used Eggleton code (Eggleton 1971, 1972). It is a Henyey-type code that uses a relaxation method to evolve the entire stellar structure. This has the advantage that mass loss occurs at the stellar surface rather than an interior point and convective dredge-up of elements to the photosphere produced by nucleosynthesis is identifiable. It simultaneously solves composition equations with structure and adaptive mesh equations. For cool star mass loss we employ Reimers mass loss prescription (Reimers 1975) augmented by an empirical fit to Mira Variables and OH/IR sources described in Lawlor & MacDonald, 2006. Briefly, this additional mass loss is calculated as a function of oscillation period where the period is taken to be function of stellar-model surface properties (Ostlie & Cox 1986). For evolution to the blue ($T_{eff} \geq 10^4$ K), we use a modified form of the mass-loss law of Abbott (1980,1982). We showed in Lawlor & MacDonald (2006) that this is in good agreement with observed mass loss rates for central stars of planetary nebula and PG1159 stars.

## 2.2 Late Thermal Pulse evolution: New models

We construct three new series of models, and we include one existing series from LM03. All models are full cradle-to-grave evolution calculations, evolved from the pre-main sequence through white dwarf cooling track. For our initial helium abundance we adopt Y = 0.249 based on abundances derived by Casagrande et al. (2007). To create a range of new LTP models, we manually vary mass loss beginning at the peak of the final AGB thermal pulse, which varies the helium shell mass as the model departs the AGB. Depending on the initial properties such as mass and metallicity, there is a range of helium layer mass that will result in an LTP after leaving the AGB, but before entering the WD cooling track. Adjusting mass loss down has the effect of delaying when the model begins to contract and leave the AGB. This occurs approximately when the hydrogen mass in the model-star is reduced by mass loss to the canonical value, $Log(M_H) = -3.0$. Thus we are adjusting the amount of helium with respect to the final AGB mass of hydrogen.

We base our new initial model choices on previous published estimations of mass for V839 Ara. Estimates have included 0.59 $M_\odot$ (Bobrowski 1994), 0.6 $M_\odot$ (Parthasarathy 2000), 0.53 $M_\odot$ to 0.56 $M_\odot$ (RNDL17), and 0.56 $M_\odot$ (Lawlor, 2019). RNDL17 et al. based their range on a comparison between $Log(g)$ and $Log(T_{eff})$ and evolution models published by Miller Bertolami (2016). They further estimate an upper limit for initial mass of 1.5 $M_\odot$ based on the observed sub-solar carbon abundance shown in the V839 Ara's spectra, and persistent solar-like abundances. The reasoning they give for this is that stars below this initial mass do not experience the AGB third dredge up (Cristallo et al. 2015; Girardi & Marigo 2007). Thus, we choose models that result in remnant masses within or close to this range of estimates when the model reaches the evolutionary point where V839 Ara is believed to be (at minimum luminosity). Also, we choose an initial mass below 1.5 $M_\odot$. Thus, we choose an initial mass of M = 1.2 $M_\odot$ as our primary model for this investigation.

Our choice for initial metallicity is Z = 0.015, adopted based on RNDL14's determination that the star has persistently had approximately solar compositions. RNDL14 note that this composition is in line with the star's physical location in the galaxy. To bracket our comparisons, we also include models from an initial M = 2.0 $M_\odot$ and 1.2 $M_\odot$ with Z = 0.030 and one series taken from LM03 with M = 1.0 $M_\odot$ and Z = 0.010. We warn that LM03 models are evolved using only Reimers mass loss and so likely overestimate the final remnant masses relative to the initial model mass and metallicity. We summarize our initial models Table 1.

Table 1 Summary of initial models used to construct LTP evolution tracks.

| Mass ($M_\odot$) | Z (mass fractions) | Remnant Mass (($M_\odot$) |
|---|---|---|
| 1.00* | 0.010 | 0.560 – 0.575 |
| 1.20 | 0.015 | 0.545 – 0.554 |
| 1.20 | 0.030 | 0.521 – 0.523 |
| 2.00 | 0.030 | 0.627 – 0.650 |

* from LM03

## 2.3 Late Thermal Pulse evolution: The general picture

Late thermal pulse models are described and identified as type III & IV helium pulses in Lawlor & MacDonald (2006). Type IV is an LTP that occurs at a temperature below $\text{Log}(T_{eff}) \approx 4.47$ and type III occurs above. The primary reason they choose this division is that an effective temperature above this is high enough to ionize a planetary nebula. In the same paper, they defined a unitless parameter that is a function of helium layer mass as a fraction of the entire helium layer mass possible, extrapolated from previous AGB thermal pulses. Put simply, when and whether an LTP occurs during post-AGB evolution is a function of helium mass remaining in the helium shell at AGB departure. LTP behaviour can also be affected by where in the hydrogen-helium pulsation cycle the model is when it reaches the end of the AGB.

In Figure 1 we show an example for a typical series of LTP evolution models taken from the LM03 series. We show that for models that originate from the same point on the AGB, the extent to which an LTP loop evolves toward the blue before turning back toward the AGB is inversely proportional to how much helium remains. The black line in Figure 1 is for a model with a helium layer mass above the mass range that leads to LTP behaviour (LTP He range – see Lawlor & MacDonald 2006)). This model experiences an additional AGB helium pulse while the mass of hydrogen is still greater than approximately 0.001 $M_\odot$, before the model leaves the AGB. For this model, the additional thermal pulse reduces the overall mass of helium to a value below the LTP He range by the time the hydrogen mass is reduced to the canonical value. Thus, the model leaves the AGB and enters the WD cooling track as a DA white dwarf with no post-AGB helium pulses. As mass loss is increased for this series and helium layer mass is reduced, this behaviour transitions to LTP behaviour. One transitional model is show in red in Figure 1. For this model the hydrogen mass is lower ($\text{Log}(m_H) = -2.63$) than for the black line and it has evolved closer to AGB departure when the helium pulse occurs.

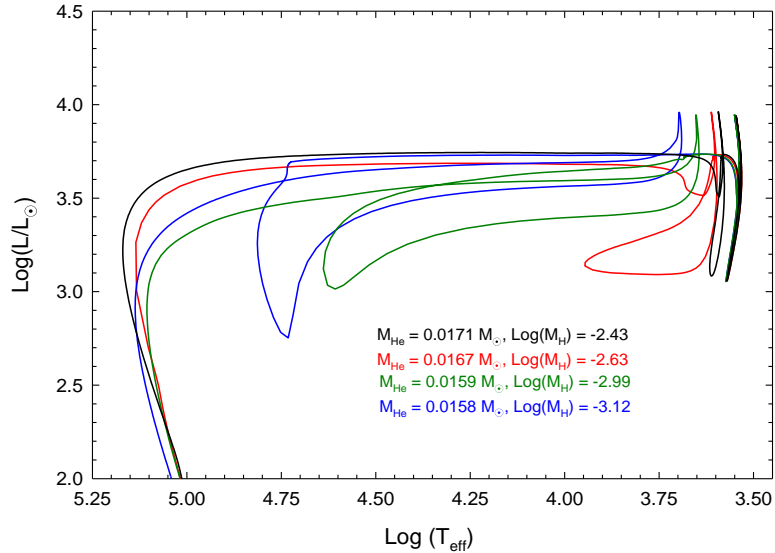

**Figure 1** A representative sample of late thermal pulse models show over a range of helium shell masses. (models from Lawlor & MacDonald 2003). The initial model for this series has M=1.0 M$_\odot$ and Z=0.010. The remnant masses range from 0.574 M$_\odot$ (blue) and 0.579 M$_\odot$ (black).

The green evolution track in Figure 1 shows a model that departs the AGB but experiences a type IV late thermal pulse, peaking at $\mathrm{Log}(T_{\mathrm{eff}}) = 3.68$. If the helium layer mass is instead at the lower end of the LTP He range while leaving the AGB, it will experience a late helium flash at a higher temperature as seen for our blue track. This model experiences a type-III LTP that peaks when the model reaches a temperature $\mathrm{Log}(T_{\mathrm{eff}}) = 4.74$.

If a model star leaves the AGB with a helium layer mass just less than the lower end of the LTP He range due to increased mass loss, it will have no LTP and enter the WD cooling track where it may experience a VLTP. If the helium layer mass is below the VLTP He range, the model will enter the WD cooling track as a DA white dwarf and experience no late helium pulse. The specific values of helium layer mass ranges depend on initial model properties, detailed by Lawlor & MacDonald (2006).

## 3. MODEL-OBSERVATION COMPARISONS

SBS21 identify a number of challenges for the LTP scenario. For one, they note that there have been no published models that can match timescale of the fast vertical evolution, decline in luminosity. They also object that no papers have addressed the sudden (decades rather millennia) ionization of the PN and they question the nebula's rapid apparent recombination. They note that the ionization occurred coincident with a mass loss episode that suddenly turns off. Another challenge is that the central star becomes less luminous and cooler than most evolution models of appropriate mass. With our models, we can identify a period during which our models do match observed timescales. We provide a solution to why V839 Ara evolves to cooler temperatures compared to (most) existing models and suggest that the mismatch for the decrease in luminosity is not alone disqualifying. Last, we propose a qualitative solution for how the nebula may have become ionized on a short timescale, and faded rapidly between 1996 and 2016.

**3.1 Observations for comparisons**

In Table 2 we present the luminosity and temperature of V839 Ara published in SE15 and we use these to calculate stellar radius and Log(g). Log(g) values are calculated using a range mass between 0.50 $M_\odot$ and 0.63 $M_\odot$. Table 2 points are for the earliest part of the observed evolution and are estimated from reported spectral types, or estimations of spectral types (SE15). We choose this range because RNDL17 have approximated the mass to be between 0.53 $M_\odot$ and 0.56 $M_\odot$, Lawlor (2019) used preliminary models to estimate a mass of 0.56 $M_\odot$, while Parthasarathy (2000) estimated a mass of 0.60 $M_\odot$.

**Table 2** Values of observational parameters we use for V839 Ara, taken from Schaefer and Edwards (2015). We have calculated Log(g) from Schaefer and Edwards approximations for a range of mass values from 0.50 $M_\odot$ to 0.63 $M_\odot$.

| | Schaefer & Edwards (2015) | | | Log(g) cm/s$^2$ | | |
|---|---|---|---|---|---|---|
| Year | Log(R) (cm) | Log(T$_{eff}$) | Log(L/L$_\odot$) | 0.50 $M_\odot$ | 0.55 $M_\odot$ | 0.60 $M_\odot$ |
| 1001 | 12.84 | 3.70 | 3.74 | 0.15 | 0.19 | 0.23 |
| 1889 | 11.28 | 4.48 | 3.74 | 3.27 | 3.31 | 3.35 |
| 1980 | 11.16 | 4.45 | 3.38 | 3.51 | 3.55 | 3.59 |

In Table 3, we present Log(T$_{eff}$) and Log(g) taken from RNDL14 and RNDL17 along with our calculations for Log(R)(cm) and Log(L/L$_\odot$). Our calculations require mass estimates, and so we present these as a range of values. Reindl et al. obtain their values by using observed data from the International Ultraviolet Explorer (IUE), the Far Ultraviolet Spectroscopic Explorer (FOS) and HST and they fit them to synthetic spectra. Table 3 also contains an approximation for 1971 from Parthasarathy (1995, 2000) and it includes the most recent HST observation taken from RNDL17. It is worth reiterating that values for luminosity can vary widely for small errors in Log(g), and here the error for Log(g) is +/− 0.50. Thus, HR Diagram comparisons are likely only useful for a comparison to general trends and timescales. The Log(g)-Log(Teff) may be better suited for comparisons, though there too the error is not insignificant. For this reason we consider a variety of comparisons, including time scales.

**Table 3** Values of observational parameters we use for V839 Ara taken from Reindl et al. (2014) for 1988 to 2015 and from Parthasarathy (2000) for 1971. Parameters presented with a range represent values we have calculated from these sources but for a range of masses, from 0.50 M$_\odot$ to 0.63 M$_\odot$.

| | Reindl et al.,(2014, 2017) | | 1971 data: Parthasarathy (2000) | |
|---|---|---|---|---|
| Year | Log(R)(cm) | Log(T$_{eff}$) | Log(L/L$_\odot$) | Log(g) cm/s$^2$ |
| 1971 | 11.67 - 11.70 | 4.322 | 3.91 - 3.95 | 2.50 |
| 1988 | 10.52 - 10.55 | 4.580 | 2.64 - 2.69 | 4.80 |
| 1992 | 10.42 - 10.45 | 4.633 | 2.65 - 2.70 | 5.00 |
| 1993 | 10.42 - 10.45 | 4.643 | 2.69 - 2.74 | 5.00 |
| 1994 | 10.32 - 10.35 | 4.681 | 2.64 - 2.70 | 5.20 |
| 1995 | 10.32 - 10.35 | 4.699 | 2.71 - 2.77 | 5.20 |
| 1996 | 10.32 - 10.35 | 4.699 | 2.71 - 2.77 | 5.20 |
| 1997 | 10.17 - 10.20 | 4.699 | 2.41 - 2.47 | 5.50 |
| 1999 | 9.92 - 9.95 | 4.740 | 2.08 - 2.13 | 6.00 |
| 2002 | 9.92 - 9.95 | 4.778 | 2.23 - 2.28 | 6.00 |
| 2006 | 9.92 - 9.95 | 4.740 | 2.08 - 2.13 | 6.00 |
| 2015 | 10.17 – 1020 | 4.699 | 2.41 - 2.47 | 5.50 |

### 3.2 Model-observation comparisons in the HR diagram and timescales

Before we address the question of high Log(g) and low Log(L/L$_\odot$) for observations compared to published models, we first note that there are a number of difficulties in making such comparisons both from the stand point of observational errors and that evolution models depend on numerous pre-set parameters that can lead to a wide range of behaviours. Among these are initial mass, initial metallicity, initial H and He compositions, convective mixing efficiency, mass loss, and at what point during the final AGB pulse mass loss is first adjusted to name some. Each of these can have an impact on the outcome of post-AGB evolution. Indeed we leverage the effect of changing mass loss in order to produce a wide range of behaviours, including LTPs.

Other difficulties arise from observational error. Values used here for observed luminosity depend on Log(g) and Log(T$_{eff}$). Log(g) values taken from RNDL14 include an error of +/– 0.50 and they suggest that a more sensitive determination may be useful. This error in gravity propagates in calculations for luminosity and radius. For example, if Log(g) is at the lower bound of error for the minimum luminosity in 2006 (5.5 instead of 6.0), the minimum luminosity for 2006 would change from Log(L/L$_\odot$) = 2.09 to 2.58, a significant increase for comparisons. Likewise the radius for an assumed mass M = 0.55 M$_\odot$ is calculated to be Log(R) = 10.17 rather 9.93.

Aggravating comparisons more, and perhaps most important, it is evident from Figure 1 that minimum luminosities for LTP model loops do not depend strongly on the model's post-AGB mass. For each model in Figure 1 and Figure 2, the masses are no more than a few percent different, but some models decline steeply while others only decline modestly in luminosity.

Instead the depth of decline depends on the temperature at which the late helium pulse occurs and this in turn depends on the remaining mass of helium when the model leaves the AGB as well as where in the hydrogen and helium pulsation cycle the model is when it reaches the end of the AGB. Lastly, the earliest observed points that we use from SE15 are approximations based on earlier reported spectral types. Thus, we conclude that identifying a 'perfect' model fit may not be realistic or necessary. However, we can show that our models' general evolution and time scales provide strong evidence that V839 Ara is indeed a late thermal pulse object, albeit a unique type (as we will show).

In Figures 2 – 4 we present four series of LTP model evolution tracks compared to observations described in section 3.1. The observations are presented as a range between red and blue triangles, where the red series represents luminosities calculated with $M = 0.50\ M_\odot$ and the blue series is calculated using $0.63\ M_\odot$. This approximates our lowest and highest model masses at the time an LTP occurs. For the three SE15 points included, the triangles are on top of each other because these were approximated from earlier spectral classifications rather than calculated using mass estimates. The remaining points are from RNDL14 and RNDL17 and a single 1971 point is estimated from Parthasarathy (1995, 2000). The models shown in Figure 2 are our preferred tracks for V839 Ara (to date), as we will justify. The initial model for this series has $M = 1.2\ M_\odot$, $Z = 0.015$, a solar composition similar to that reported for V839 Ara by RNDL14. Although the decline in model luminosity does not reach as deeply as the observations, they do reach the upper limit of the error bars and as we argued, the depth of model LTP luminosity does not depend on primarily on mass and should not in itself disqualify it. This series does produced one model (shown in the right panel of Figure 6) that reaches closer ($\text{Log}(L/L_\odot) \approx 2.3$), but the LTP erupts at too high a temperature ($\text{Log}(T_\text{eff}) = 5.0$) to explain V839 Ara. The two models shown with mass $0.553\ M_\odot$ and $0.547\ M_\odot$ in Figure 2 have the advantage that they decline in luminosity beginning at a temperature comparable to that at which V839 Ara likely began to decline, $\text{Log}(T_\text{eff}) \approx 4.65$ to $4.75$. In these and all LTP models, the helium flash reaches its peak approximately concurrent with the sharp turn downward in luminosity.

Aside from declining at a comparable temperature, our preferred models as well as the $Z = 0.01$, $M = 0.575\ M_\odot$ model from LM03 shown in Figure 3, exhibit a unique 'knee' feature, or inflection, as they begin to decline vertically. For a very brief period the model cools before declining vertically in luminosity. This feature is highlighted by the black evolution track ($M = 0.553\ M_\odot$) and as we will show, may be specific to the type of LTP that V839 Ara has (likely) undergone. This 'knee' feature always occurs in this temperature range in these and other model series' not yet published (nor pertinent to V839 Ara). On these grounds, we cannot rule out the LM03 model, shown in Figure 3, however we note that the LM03 model was constructed prior to employing our more recent mass loss scheme and its mass is likely overestimated.

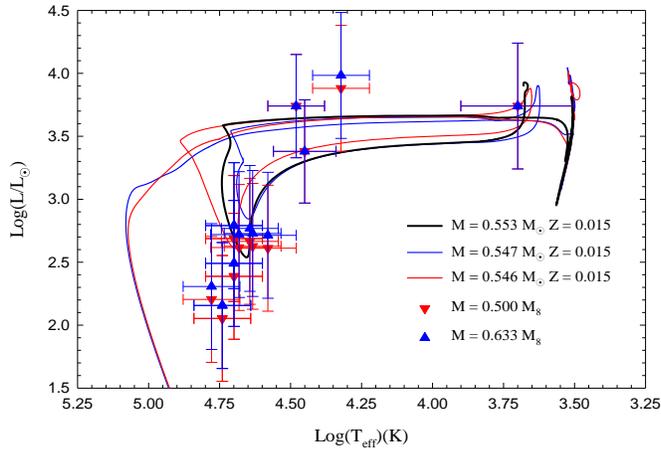

**Figure 2** Three 'preferred' models evolved from our series with M = 1.2 M☉ and Z = 0.015. Remnant masses are shown for each line, and the blue and red triangles correspond to V839 Ara observations at upper and lower mass limits. Luminosities for the red triangles are calculated using M = 0.50 M☉ while blue triangles are calculated using M = 0.63 M☉.

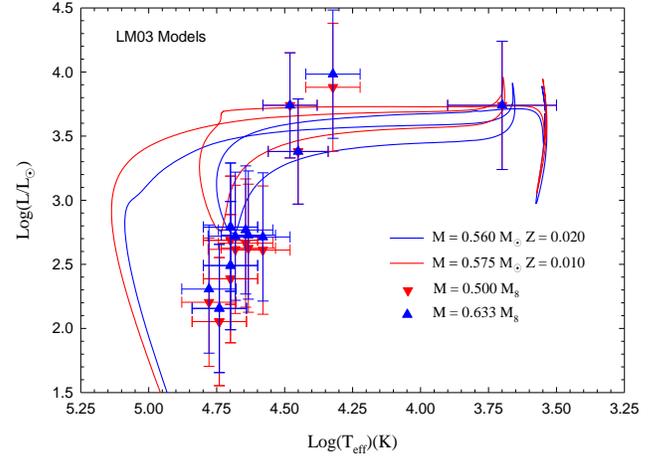

**Figure 3** Two models evolved from our LM03 series with M = 1.0 M☉, Z = 0.010 (red) and Z = 0.020 (blue). Remnant masses are shown for each line and red and blue triangles are the same as for Figure 2

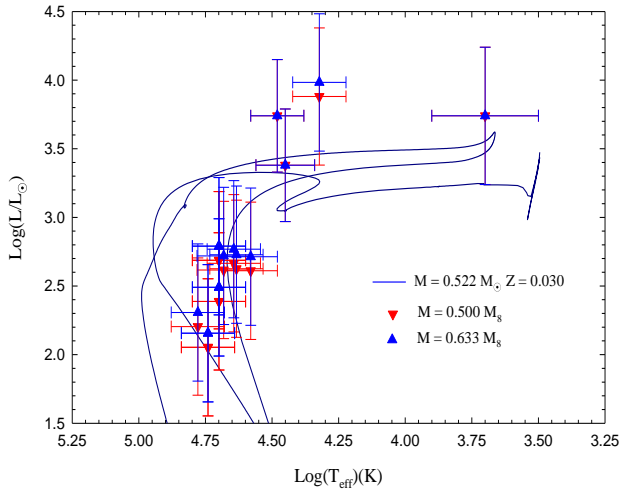

**Figure 4** One model evolved from our series with M = 1.2 M☉ and Z = 0.030. Remnant mass is shown (dark blue), and the red and blue triangles are the same as for Figure 2.

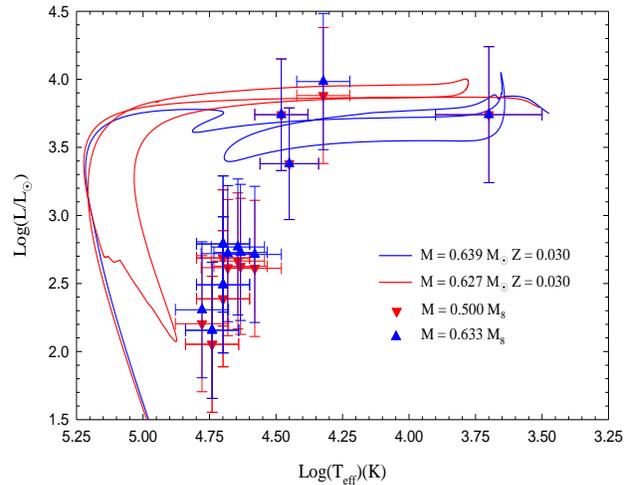

**Figure 5** Two models evolved from our series with M = 2.0 M☉ and Z = 0.030. Remnant masses shown for each line and the red and blue triangles are the same as for Figure 2.

We are convinced that the final two model series shown in Figures 4 and 5 can be ruled out except for comparison's sake and potential future discoveries. In Figure 4 we show a model with initial M = 1.2 M☉ and Z = 0.030. This model has some features that are similar to V839 Ara, such as a deep decline in luminosity; however, the departure from the AGB and crossing to the PN phase is under luminous. Luminosity during this time is controlled by mass, and by the time the model leaves the AGB, it has a low mass (0.522 M☉) due to a high mass loss driven by high surface opacity (Z = 0.030). Based on this model, we adopt a lower limit for mass of higher than this (0.54 M☉). Another strike against this series is that nearly all the LTP models it produces occur either at a temperature that is too high or too low. Some from this series exhibit the unusual behaviour of experiencing both, but none decline at a temperature similar to V839 Ara. Last, in Figure 5 we compare models evolved from an initial mass M = 2.0 M$_\odot$ and Z = 0.030. This model reaches the minimum luminosity of V839 Ara easily, but at too high a temperature. This series too only experiences LTP's that ignite when its temperature is too hot or too cool. Thus we rule this model out. Though searching for a model of this mass that experiences an LTP at an appropriate temperature provides a tempting solution, no other published estimations have found this high mass to be likely.

Another challenge posed for the LTP scenario is that published LTP timescales are slow compared to the vertical decline in observed luminosity, only about twenty years (SE15). While it is true that for our models the LTP takes several hundred years for the pulse to ignite and reach its peak helium burning luminosity, the time scales from the peak of helium burning luminosity to the point where the overall stellar luminosity reaches its minimum value (point B in Figure 6) is on the order of a few decades or less. This point corresponds to evolutionary state of V839 Ara in 2006. We show in Figure 6 two models from our M = 1.2 M$_\odot$, Z = 0.015 series in the HR diagram with evolutionary points labelled in years. This figure shows an LTP that occurs at a temperature close to the temperature at which V839 Ara began to decline in luminosity (left) and a type III LTP that erupts at a higher temperature (right). In both cases the decline in stellar luminosity occurs on a very similar timescale as the decline of V839 Ara (approximated in Table 4). In both Figure 6 and Table 4 we show time scales in years between four points: point A is the peak of helium burning for the LTP, point B is the minimum stellar luminosity, point C and D are both subsequent increases in Log(L/L$_\odot$) of 0.25 each. In Table 4 we include timescales for four additional models from the same series during the same corresponding periods. Nearly every LTP model provides a reasonable timescale match, except our 0.554 M$_\odot$ model which is an early type IV LTP, erupting at a cool temperature (Log(Teff) = 3.63) and remaining comparatively bright. For these six sample models, two experience an LTP at a temperature that corresponds to the temperature similar to that at which V839 Ara began its rapid decline in luminosity. We will show that this specific type of model is the best type-fit for V839 Ara. Also in Figure 6 we show the time scale for the track to traverse back to the AGB. In all cases this happens on a time scale of a few hundred years – a similar timescale to what is assumed for FG Sge (Lawlor & MacDonald 2003, Gonzalez et al 1998).

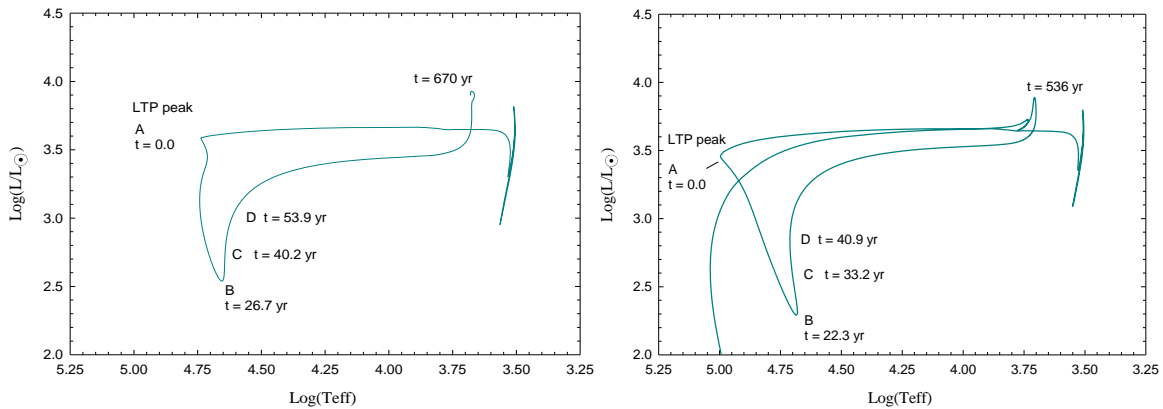

**Figure 6** Two LTP models with evolution time scales labeled beginning form the LTP helium burning peak, as shown. The 0.553 M$_\odot$ model in the left panel is a 'V839 Ara' type LTP (a near match for temperature at which its luminosity declines) and the panel on the right is for a 0.546 M$_\odot$ type III LTP, which erupts near Log(T$_{eff}$) = 4.9.

**Table 4** Time scales for our 1.2 M$_\odot$, Z = 0.015 model series between points labeled in Figure 6. Also shown is the temperature at which the LTP reaches a peak value.

| Model/Object (M$_\odot$) | LTP Log(T$_{eff}$)(K) | $\Delta t_{(A-B)}$ (yr) | $\Delta t_{(B-C)}$ (yr) | $\Delta t_{(C-D)}$ (yr) |
|---|---|---|---|---|
| 0.554 | 3.63 | 84.8 | — | — |
| 0.553 | 4.73 | 26.7 | 13.6 | 13.5 |
| 0.552 | 4.97 | 22.3 | 10.9 | 7.7 |
| 0.550 | 4.89 | 30.6 | 15.8 | 15.9 |
| 0.547 | 4.70 | 20.6 | 17.7 | 23.6 |
| 0.546 | 4.86 | 22.1 | 13.6 | 16.9 |
| V839 Ara | ≈ 4.70 | 22 | 11.0 | — |

Because of the rapid evolution of LTP's, we can test our timescales in the relative near future because the time between points C and D is also very short (10 – 20 years, see Table 4). In fact, one test has possibly already passed: In Table 4 we have estimated the timescale corresponding to the time between points B and C for the central star V839 Ara, assuming that the luminosity in 2006 was indeed the minimum (RNDL17). RNDL17 reported that for the star Log(L/L$_\odot$) increased by 0.25 between 2006 and 2015. This is in quite good agreement with our models between corresponding points B and C. We note here though that SBS21 were not sure that luminosity was indeed increasing.

Thus, our model timescales do predict the rapid decline of V839 Ara. To strengthen our timescale argument further, we note that the time scale for our model leading up the LTP peak and the sudden luminosity decline is also in line with observations. Between 1889 and the early 1990s, observations show a horizontal increase in temperature from about Log(T$_{eff}$) = 4.5 to 4.6, occurring on the order of 100 years. For our M = 0.547 M$_\odot$ model, the timescale for a comparable change in temperature is 120 years. We take this time from our model LTP peak at Log(Teff) = 4.7 back to a point when the model has Log(T$_{eff}$) = 4.6. Further, the rate of temperature evolution from red to blue is relatively constant back to at least Log(T$_{eff}$) = 4.0. Finally, SE15 roughly approximate 1000 years between V839 Ara's AGB departure and the appearance of the nebula (Log(T$_{eff}$) = 3.7 to 4.48). Both of our preferred models traverse this in approximately 4500 years, however they both evolve from the top of SE15's error bar in temperature (Log(T$_{eff}$) = 3.9) in about 900 years.

### 3.3 Comparisons in the Log(g)-Log($T_{eff}$) plane

We make comparisons in Figure 7 in the Log(g)-Log($T_{eff}$) plane. While using such diagrams is an effective way to estimate masses for PN and related objects, LTP object comparisons suffer the same shortcoming as HR diagrams. That is, models of effectively the same mass can reach very different values of Log(g). Our lowest mass 0.522 $M_\odot$ model shown in the bottom left panel reaches near the highest gravity. Both this and our 2.0 $M_\odot$ model (bottom right) have an LTP that erupts at a temperature too high, or too low in the case of the 0.650 $M_\odot$ remnant (blue line, bottom right). Thus here again, we focus our comparisons on temperature more so than remnant mass.

For our preferred models (Figure 7, top left panel) we see that only our 0.546 $M_\odot$ model (black line) attains a Log(g) ≈ 6.0 however its temperature is too high. The point of maximum gravity in that model corresponds to minimum radius and is nearly coincident with the peak helium burning luminosity for the LTP which is the point when model evolution begins to accelerate. It appears that models all suffer from temperatures that are too high at maximum gravity compared to observations. However; there is something more happening here for V839 Ara, and our preferred models offer a clue. The models in Figure 7 all increase in gravity along a very narrow corridor during early contraction, so that all masses can explain the early observations and have an appropriate slope. Virtually no models predict or evolve directly along the line of the most recent observed values for Log(g) from RNDL14. In fact, on careful inspection it even appears that the most recent observations are shifted out of line (i.e.; cooler) with the earliest observed points.

We believe that this *is not* due to error or coincidence. In Figure 8 we show our preferred models in an expanded view of the Log(g)-Log($T_{eff}$) plane, comparing just the observations from 1971 to 2015. Although our models do not evolve to as high a value for gravity as the observations (though they are close to the upper limits), they do show a conspicuous *sharp turn cooler* before evolving to the maximum value for Log(g) (see the thick red and blue evolution tracks). Still, why is observed gravity higher then models? One possibility is that perhaps a dynamic mass loss episode coincident with the flash (see section 3.4), unable to be modelled here, could lead to a smaller and higher gravity remnant. This behaviour in the Log(g) – Log($T_{eff}$) is exactly concurrent with the 'knee' feature we highlighted in the HR diagram, and it occurs just after the LTP peak and just before the rapid (≈20 years) decline in luminosity. This feature provides an explanation for why the most recent observations from RNDL14 and RNDL17 for Log(g) have eluded a good match to many published model calculations, and appear unusually cool in comparison.

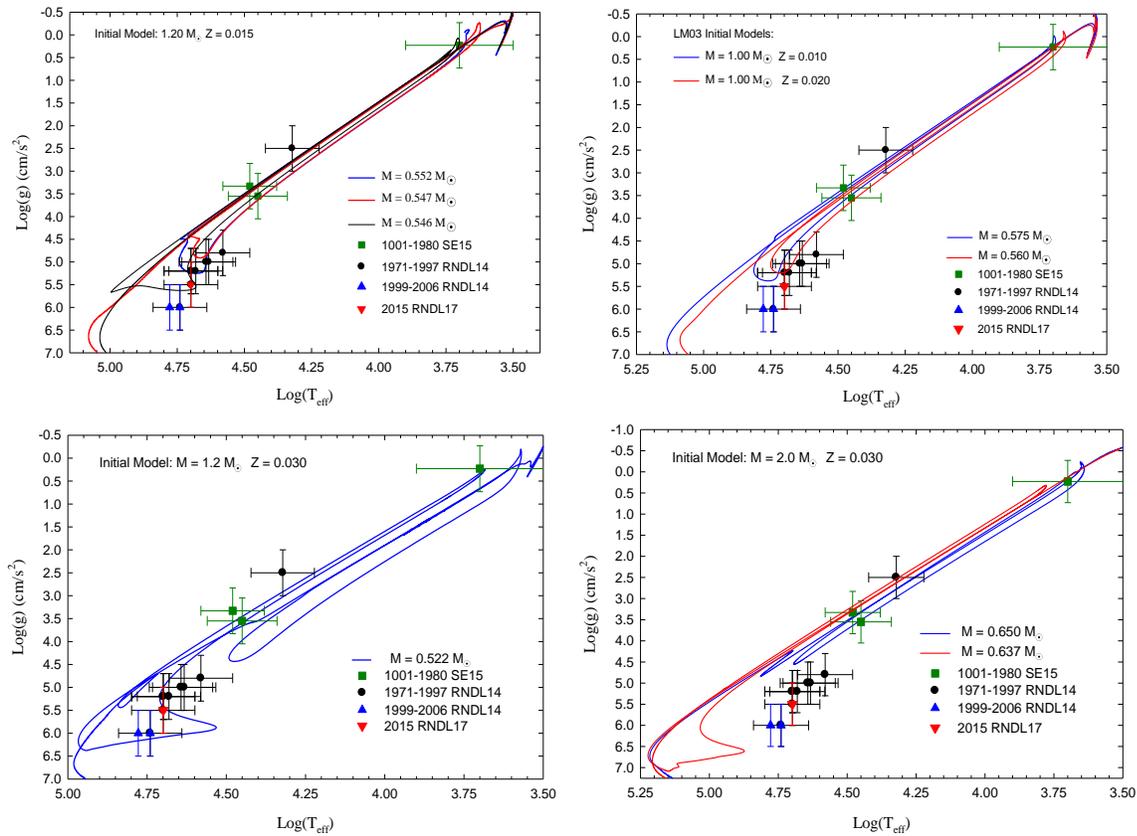

**Figure 7** Log(g) - Log(Teff) for four series of LTP models compared to V839 Ara taken from Reindl et al. (2014, 2017), Schaefer et al. (2015), and Parthasarathy, (2000).

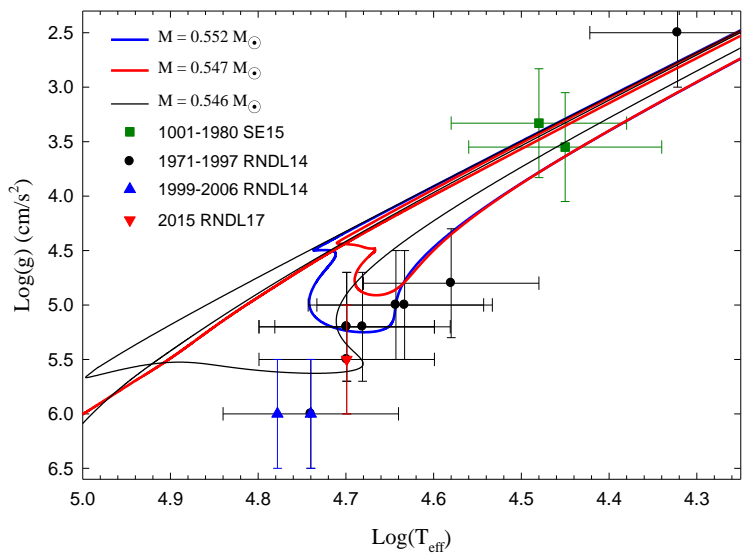

**Figure 8** Comparisons between our M = 1.2 M☉ Z = 0.015 series of LTPs and V839 Ara showing a sharp move cool before reaching maximum Log(g) for those that experience a secondary hydrogen flash (red and blue lines).

To explain this brief sharp turn toward cooler temperatures, we present in Figure 9 the time evolution of the hydrogen and helium burning luminosity for three models just before and after the helium burning (LTP) peak. The top panel in this figure is for our model with M = 0.547 M$_\odot$, more specifically one for which the LTP erupts between Log($T_{eff}$) = 4.6 and 4.7 and peaks between Log($T_{eff}$) = 4.7 and 4.8. This is the model the exhibits the 'knee' feature in the HR diagram. The middle and bottom panel of Figure 9 are also for LTP models with slightly different masses evolved from the same model series. The clear difference is that in the middle and bottom panels, hydrogen burning luminosity has declined further than it has in the top panel by the time that the LTP peaks. For our M = 0.546 M$_\odot$ in the middle panel, the LPT occurs at Log($T_{eff}$) = 4.86 and the subsequent hydrogen pulse is less intense. In the bottom panel we show our M = 0.554 M$_\odot$ model for which the hydrogen fusion declines significantly more, and by the time the model experiences an LTP, hydrogen fusion is on its way to being extinguished. The LTP peak for this model occurs when Log($T_{eff}$) = 3.63. Thus, models that experience an LTP in a temperature range corresponding to that at which V839 Ara suddenly declined are distinguished by a significant ignition of hydrogen due to convective mixing (we describe this in the following section). During this time, hydrogen burning luminosity increases to 1.83 ×10$^6$ L$_\odot$ for the model shown.

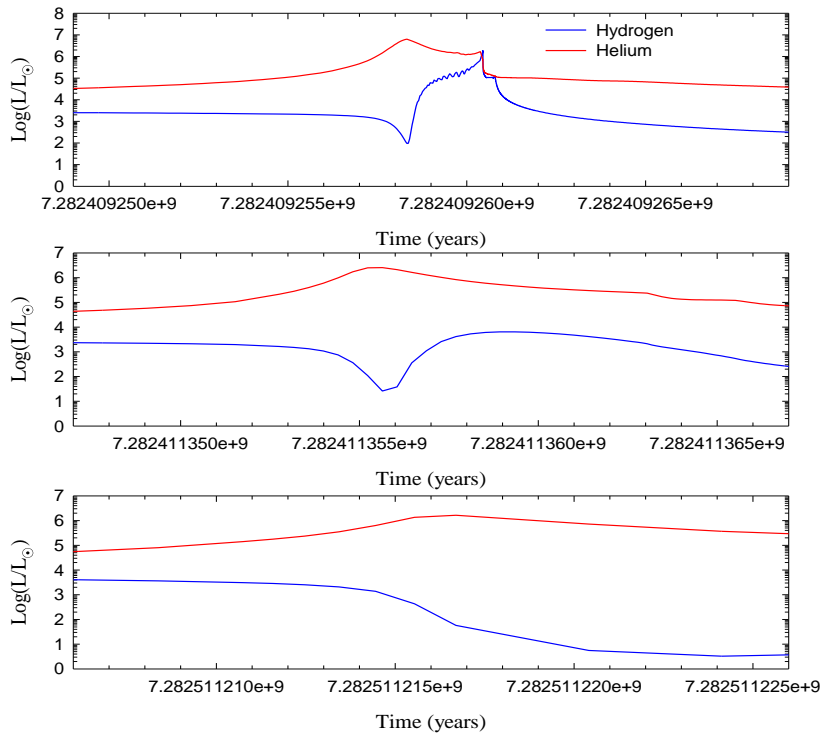

**Figure 9** Hydrogen burning luminosity (blue lines) and helium burning luminosity (red lines) for three models. The top panel is for a 0.547 M$_\odot$ model, the middle panel is for a 0.546 M$_\odot$ model and the bottom panel is for a 0.554 M$_\odot$ model. For the top panel the LTP peak occurs at Log($T_{eff}$) = 4.70, a temperature that corresponds approximately to temperature at which V839 Ara declines in luminosity. For the middle panel, the LTP peak occurs at a higher temperature, Log($T_{eff}$) = 4.86 and for the bottom panel the LTP peak occurs at a cooler temperature, Log($T_{eff}$) = 4.63.

To summarize, for models created with the lowest mass loss, reduced by more than 50%, AGB departure is delayed and hydrogen fusion declines the most (bottom panel of Figure 9). For models created with higher mass loss, reduced by less than 25%, the model evolves more quickly and nearly reaches the PN phase by the time the LTP erupts (middle panel). These models experienced a moderate secondary hydrogen pulse on the order of $10^3 L_\odot - 10^4 L_\odot$ before cooling and declining in luminosity. Models evolved with intermediate mass loss, reduced between 30% and 50%, may result in a secondary hydrogen flash ($10^5 L_\odot - 10^6 L_\odot$). Thus we conclude that V839 Ara underwent a specific variety of LTP that includes a secondary mixing induced hydrogen flash close to the surface, evident in its move cooler in the $Log(g)$-$Log(T_{eff})$ plane. This provides a clue to the last difficulties with the LTP scenario.

### 3.4 The ionization of the Stingray Nebula and 1980 – 1993 mass loss event

There has been little published investigating the impact of late thermal pulses on a star's surrounding nebula. Early on, Tylenda (1979) modelled nebulae including post-AGB helium flash effects and reported that time-dependent effects are expected to occur in ionization-bounded nebulae of moderate and high excitation during flashes, and in the same paper suggest that helium flashes can ionize the nebula. Tylenda's models showed that a helium flash can account for OI in the inner shell and OII and NII in the outer shell of a nebula. Earlier, Paczynski (1971) reported that if the mass of a hydrogen envelope is small, the effective temperature can vary drastically during a helium flash, enough to ionize the nebula. Specific to the Stingray Nebula, SBS21 specified "…the stark coincidence in time between the 1980s ionization event and the 1980s mass loss event carries a strong case that the two are causally related." They point out that mass loss was close to $10^{-9}$ $M_\odot$/year from 1988-1993, and then dropped by ten times in 1994.

We cannot model dynamic effects of an LTP on temperature and mass loss due to our code's requirement of hydrostatic equilibrium. However, we can show the condition described by Paczynski does exist at and near the peak of our LTP models. In Figure 10 and Figure 11 we provide a qualitative solution for this seemingly intractable problem. In Figure 10 we show the hydrogen and helium profiles and the nuclear energy generation in solar luminosities both as a function of stellar mass for our M = 0.547 $M_\odot$ model. Both are shown at a time shortly following peak helium and hydrogen burning luminosity. An important feature of these figures is that we are only showing the structure for the outer 10% of the shrinking star and that the regions where the most energetic fusion occurs are in the outer 5%. The late helium pulse is visible in the top panel of Figure 10 near 0.52 $M_\odot$ and there is significant energy generation due to mixing and consumption of hydrogen visible near 0.54 $M_\odot$. The mixing here is due to a convection zone that reaches from 0.535 $M_\odot$ to just beyond the helium-hydrogen shell boundary. This much hydrogen consumption is not a feature in all LTP models, but is for our preferred models for V839 Ara. This is the source of the sharp right turn redward noted in the $Log(g)$-$Log(T_{eff})$ plane in Figure 8.

We note that the hydrogen envelope at this point is very thin, comprised of less than the outer 0.3% of the structure. We argue that it is the helium flash and secondary hydrogen consumption that is responsible for the ionization of the nebula in the 1980's. The rise in energy generation due to hydrogen consumption is extremely rapid. The mixing induced increase in hydrogen burning luminosity from $10^2$ $L_\odot$ to $10^6$ $L_\odot$ *occurs in 2.05 years* and nearly reaches the surface. The increase begins immediately after helium burning luminosity reaches its peak. By the time hydrogen burning luminosity reaches its peak value, helium burning luminosity is reduced from $6.3 \times 10^6$ $L_\odot$ to $6.7 \times 10^5 L_\odot$. Following this, the overall stellar luminosity continues to decline to its minimum value in 18.5 years. By then hydrogen burning luminosity has dropped to $10^2$ $L_\odot$ and helium is reduced to $10^4$ $L_\odot$. Because of this rapid turn off of hydrogen energy generation close to the surface, this scenario can also explain the more recent decline in brightness and the recombination of the nebula described by SE15 and Balick et al. (2021).

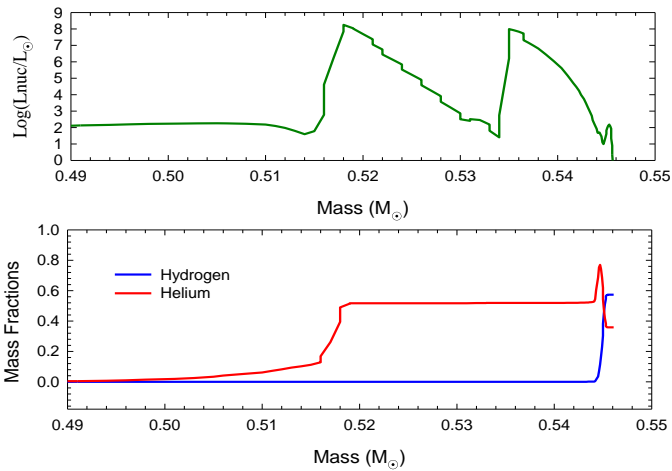

**Figure 10** Total nuclear energy generation rate (top) and hydrogen and helium compositions in mass fractions (bottom) as a function of mass (stellar structure) both at a time that corresponds to the peak of the LTP (7.282409261 ×$10^9$ yr) for our M = 0.547 $M_\odot$ model.

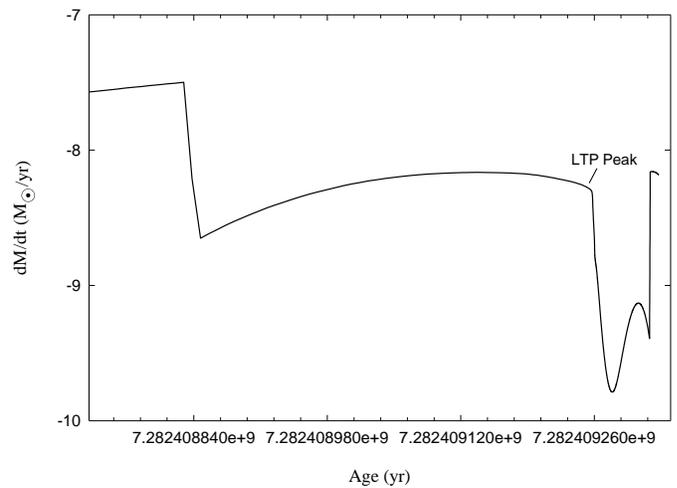

**Figure 11** LTP mass loss as a function of time for our M = 0.547 $M_\odot$ model just before and after the late thermal pulse erupts (peak labeled).

In Figure 11 we show the mass loss of a preferred model as a function of time during the increase in helium luminosity until its peak. The peak is followed by a rapid decline in mass loss. As was observed for V839 Ara (SBS21), Figure 11 shows that our mass loss increased to a few times $10^{-9}$ $M_\odot$/yr until reaching the peak of the LTP, and then declines rapidly until it drops to nearly $10^{-10}$ $M_\odot$/year. Admittedly our models are constructed using a hydrostatic equilibrium assumption thus we cannot model sudden mass loss events as described by SBS21. Additionally, RNDL17 found that the upper limit for mass loss to be $10^{-12}$ $M_\odot$/yr in most recent years, lower than for our model. This may not be surprising because the model we show in Figure 11 is for a model that does not decline in luminosity as much as the reported observations and our hot mass loss scheme depends on luminosity. Never the less, our models do predict a sudden drop in mass loss at the right time and with the right timescale.

## 4. SUMMARY AND DISCUSSION

We have created a grid of LTP models to compare to V839 Ara, the central star of the Stingray Nebula. We used a range of masses and metallicities based on previous estimations and higher mass and metallicity models for comparisons. We have answered most of the remaining doubts about the LTP scenario for V839 Ara. We have shown that the LTP scenario is indeed the most likely explanation for the behaviour of V839 Ara, and that this can be confirmed within the next decade. The timescales for our LTP model evolution match remarkably well. The sudden decline in luminosity observed for the central star began sometime in the 1980's and reached a minimum in 2006 (RNDL17), approximately 20 – 30 years. For most of our type III LTP models this decline occurs at a very comparable timescale (see Table 4). We also found that the timescale leading up to the decline in luminosity (concurrent with the LTP peak) is similar to those observed for V839 Ara over a comparable range of increasing temperature – approximately 120 years for each change in $Log(T_{eff})$ of 0.1. Further RNDL17 reported that the central star has cooled since 2006 and its corresponding luminosity then increased from $Log(L/L_\odot) = 2.08$ to 2.41 in 2015. The eleven year timescale is in very good agreement with the predictions of our models also presented in Table 4.

In our model-observation comparisons, stellar mass cannot easily be estimated solely based on Log(g) because multiple models of effectively the same mass can result in very different values of gravity depending on what temperature the LTP erupts. This in turn depends both on how much helium and hydrogen remains as the model star departs the AGB, and it depends on at what point it leaves the AGB relative to the alternating hydrogen and helium thermal pulses. For the right combination, the LTP can consume significant hydrogen if the hydrogen burning (and mass) has not declined too much by the time the LTP ignites (see Figure 9). For our 1.2 $M_\odot$, $Z = 0.015$ series of models this behaviour occurs between $Log(T_{eff}) = 4.6$ and 4.8 and when the radius has reached approximately $Log(R)(cm) = 10.68$. For example, in Table 4 our 0.553 $M_\odot$ and 0.547 $M_\odot$ models peak at $Log(T_{eff}) = 4.73$ and 4.70 respectively, while all others occur at cooler or hotter temperatures and larger and smaller radii. This is also the temperature range in which V839 Ara suddenly began to dim. We believe that this range of temperatures, radii and hydrogen masses may be roughly canonical values for such behaviour, because there is similar behaviour for the $Z = 0.01$ LM03 model, and we see it for another yet to be published low metal model. Both of these show the hallmark 'knee' in the HR diagram just before declining rapidly in luminosity at $Log(T_{eff}) = 4.73$ and 4.68 respectively. By the Lawlor & MacDonald (2006) designation, these are type III late thermal pulse models, however they are unique – type III LTP models that occur at temperatures above this range do not burn hydrogen (as intensely), nor do type IV LTP models (those that occur at less than $Log(T_{eff}) = 4.47$). Thus those that consume significant hydrogen appear to be an intermediate or transition type (type III.5, or V839 Ara type).

A conspicuous feature of the 'knee' in the HR diagram is that shortly after the hydrogen burning peak the model evolves sharply cooler in the Log(g)-Log(Teff) plane. This only occurs in models that can be described at this type III.5. The move cooler due to the addition of a hydrogen energy production provides an explanation for the apparent temperature mismatch between observations of Log(g) for V839 Ara after the 1980's and those before. Thus, not only is V839 Ara a rare observed example of an out bursting LTP, it represents a unique type that occurs between type III and type IV. This explains why the observed values of gravity have been difficult to reproduce, and why they are seemingly too cool for many LTP models that seemed to be near matches otherwise.

As early as 1971, Paczynski investigated the impact of late helium pulses on evolution of post-AGB stars and found that if a star's hydrogen envelope is thin enough it could create conditions hot enough to ionize the nebula, also argued by Tylenda (1979) to explain time-dependent effects. Discussed in section 3.4, we showed that these conditions exist in our preferred models. The hydrogen layer at the time of the LTP consists of only the outer 0.30% of the stellar structure. The LTP/hydrogen flash scenario can explain the sudden ionization of the planetary nebula coincident with the rising peak of the LTP followed by the sudden mixing induced ignition (2.05 years) of hydrogen close to the stellar surface. The overall stellar luminosity declines rapidly following the LTP peak and rapid increase of hydrogen fusion. This then explains the decline in nebular brightness from 1996 and 2016 as both the hydrogen and helium burning decline on a time scale of decades. Because Abbott's formula for hot star winds depends on luminosity, this decline also marks a sudden and precipitous drop in mass loss – also observed for V839 Ara (SBS21).

Although our preferred models do not reach values of gravity as high as observations or luminosities as low, we argue that this is not disqualifying. First, our models reach close to the limit of the error bars. If the lower limit of gravity is used, the luminosities calculated from observed gravities and uncertain masses are nearly identical to our preferred models' luminosities. We speculate that an alternative explanation may be that the 'real' star experienced a significant mass loss episode concurrent with the LTP/hydrogen flash that produced a lower mass higher gravity remnant. This is something we do not model. Secondly, we find that the depth to which models evolve in luminosity is not strongly correlated with stellar remnant mass. Instead it is sensitive to the temperature at which an LTP occurs, and this depends on a combination of many parameters including the combination of helium and hydrogen masses, and at what point the model leaves the AGB in the thermal pulse cycle. In other words, models of nearly the same mass can reach different minimum LTP luminosities depending on the temperature when the LTP occurs.

Based on our comparisons to V839 Ara we estimate by some extraction that its initial mass was between 1.0 $M_\odot$ and 1.3 $M_\odot$. We approximate its current mass to be between 0.54 $M_\odot$ and 0.56 $M_\odot$ in close agreement with RNDL17. Our preferred model from this work is 1.2 $M_\odot$ and Z = 0.015. Our comparison 2.0 $M_\odot$, Z= 0.030 model becomes too hot and our 1.2 $M_\odot$, Z= 0.030 loses too much mass and departs the AGB at a lower luminosity. The expectation for the future of this star is that it will continue to brighten on a time scale of decades, reaching as high as Log(L/L$_\odot$) = 3.25 within 20 years.

What is left is to continue monitoring both the star and nebula and determine if it fits in as an evolutionary link between other objects and PN. Questions remain: What composition changes if any will it undergo? When might it become obscured by dust as it grows and cools? What can we make of the SBS21 observations that seem to contradict the conclusion of RNDL17 that the star is cooling and brightening? Finally, can it shed light on a conclusive determination of the nature of FG Sge and related objects? We should and plan to continue the time consuming task of model calculations using parameters adjacent to those presented here. Though we have identified a mechanism to predict the cooler observations of Log(g), why exactly do models not reach higher gravities? Is this a modelling limitation or are values of Log(g) overestimated? Could a real world sudden dynamic mass loss episode, not modelled here, lead to higher values? In any case, the solution to Stingray puzzle is nearly complete.

**Acknowledgements** The author would like to thank Brad Schaefer for useful virtual discussions during AAS237 and questions that helped shape the direction of this investigation, and thanks as always to Jim MacDonald who provides invaluable modelling advice. Much of this work was supported by Penn State Brandywine's generous Cooper Fellowship Award.

**Data Availability**

*No new data were generated or analysed in support of this research.*